\documentclass[english,aps,manuscript]{revtex4}
\usepackage[T1]{fontenc}
\usepackage[latin1]{inputenc}
\usepackage{babel}
\usepackage{graphics}

\makeatletter

\providecommand{\LyX}{L\kern-.1667em\lower.25em\hbox{Y}\kern-.125emX\@}

\makeatother
\begin{document}
\hspace*{10cm} COLO-HEP-4\\
 \hspace*{10.5cm} January 2002.

{\centering \textbf{\Large Closed String Tachyons and Semi-classical
Instabilities}\Large \par}

{\centering \vspace{0.3cm}\par}

{\centering {\large S. P. de Alwis\( ^{\dagger } \) and A.T. Flournoy\( ^{\ddagger } \) }\large \par}

{\centering \vspace{0.3cm}\par}

{\centering Physics Department, University of Colorado, \\
 Boulder, CO 80309 USA\par}

{\centering \vspace{0.3cm}\par}

{\centering \textbf{Abstract}\par}

{\centering \vspace{0.3cm}\par}

{\centering We conjecture that the end point of bulk closed string
tachyon decay at any non-zero coupling, is the annihilation of space
time by Witten's bubble of nothing, resulting in a topological phase
of the theory. In support of this we present a variety of situations
in which there is a correspondence between the existence of perturbative
tachyons in one regime and the semi-classical annihilation of space-time.
Our discussion will include many recently investigated scenarios in
string theory including Scherk-Schwarz compactifications, Melvin magnetic
backgrounds, and noncompact orbifolds. We use this conjecture to investigate
a possible web of dualities relating the eleven-dimensional Fabinger-Horava
background with nonsupersymmetric string theories. Along the way we
point out where our conjecture resolves some of the puzzles associated
with bulk closed string tachyon condensation. \par}

\vspace{0.3cm}


\vfill

\( ^{\dagger } \) {\small e-mail: dealwis@pizero.colorado.edu}{\small \par}

\( ^{\ddagger } \) {\small e-mail: flournoy@pizero.colorado.edu}{\small \par}

\eject

\tableofcontents{}

\section{Introduction and the conjecture}

One of the most satisfying recent results in string theory involves
the fate of theories with open string tachyons. Sen conjectured that
the condensation of these tachyons corresponds to the decay of unstable
\( D \)-brane configurations leaving a supersymmetric vacuum (with
or without stable \( D \)-branes)\cite{Sen:1998sm, Sen:1998ii, Sen:1998rg}.
Support for this conjecture has come from techniques using conformal
field theory, open string field theory, and non-commutative geometry. 

There is in addition a natural picture that emerges wherein the open
string tachyonic instability is the perturbative manifestation (for
certain values of the background moduli) of an instability which also
admits a semi-classical description (for other values of the background
moduli). The simplest example of this type is the \( D \)-\( \overline{{D}} \)
system where for small enough separation there exists an open string
tachyonic mode which mediates annihilation of the pair a la Sen, while
for large separation the tachyonic mode becomes massive and the same
annihilation instability may be described by a sphaleron solution
of the Born-Infeld action \cite{Callan:1998kz, Savvidy:1998xx}.

\begin{figure}[!h]
{\centering \includegraphics{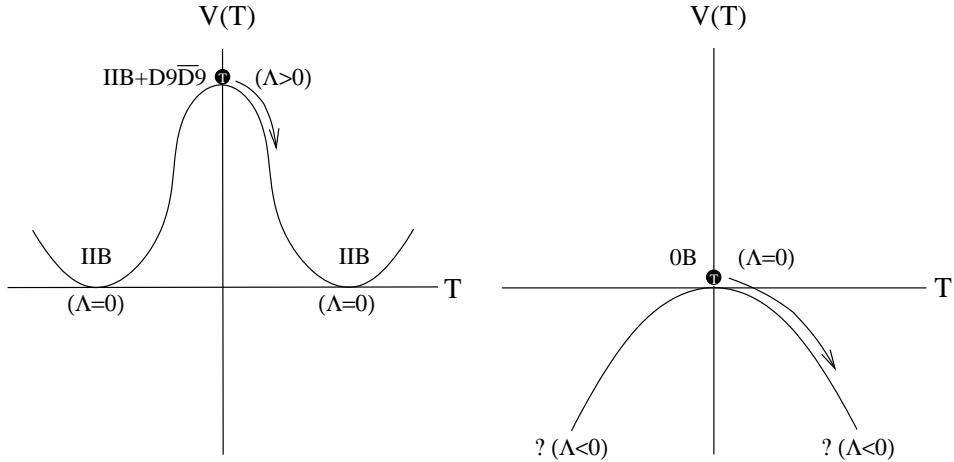} \par}

\caption{The tachyon potentials for an unstable D-brane in dS space on the
left versus the flat space tachyon of Type 0A on the right.\label{fig:SHANTA}}
\end{figure}

The case for closed string tachyons is much less understood. Recent
works on systems with closed string tachyons \cite{Gutperle:2001mb, Adams:2001sv, David:2001vm}
have all pointed towards the conclusion that the endpoint of the instability
is a supersymmetric closed string vacuum much like the case for open
string tachyons.\footnote{%
Exceptional cases include the analysis \cite{Horowitz:2000gn} of
the closed string tachyon in the purely bosonic theory and the analysis
\cite{Barbon:2001di} of the thermal tachyon associated with the Hagedorn
transition. Though the authors of \cite{Horowitz:2000gn} make no
explicit conjecture for the fate of the theory after tachyon condensation,
the implications of their discussion are along the lines presented
in this paper. In \cite{Barbon:2001di} the authors present an argument
for the decay of Wick-rotated Type 0A into supersymmetric IIB. They
first localize the thermal tachyon by imposing an AdS background geometry,
then demonstrate that the endpoint of tachyon condensation should
involve a large AdS black hole (as large as the tachyonic region)
with the spacetime behind the horizon excluded. The theory at the
transverse boundary of the orginal spacetime is then identified as
Euclidean Type IIB by a simple spin structure argument. Though this
conclusion is reminescent of what we will discuss in this paper, the
interpretation of the final state of the theory as IIB on a boundary
is not entirely clear to us. 
} However not all tachyons are equal. We believe that if the coupling
is non-zero, closed string tachyons will have a more drastic effect
on the theory than the open string tachyons for the following reason.
For open string string tachyons to arise one must have \( D \)-branes
in some closed string background (space time). According to Sen the
height of the tachyon potential is given by the tension of the relevant
\( D \)-brane(s). At the location of the decaying \( D \)-brane(s)
one has (before the decay happens and for non-zero coupling) positive
curvature. (In the case of say \( D9 \)-\( \overline{D9} \) in a
IIB background one would have dS space to begin with). After the decay
one would have a flat closed string background. On the other hand
the bulk closed string tachyon of Type 0B exists already in a flat
background. This means that the flat space background corresponds
to the unstable point (a maximum or a saddle point) of the tachyon
potential. Even if there is a minimum to the tachyon potential the
end point of the decay will not be one of the known stable flat space
backgrounds of string theory. This is illustrated in figure \ref{fig:SHANTA}.

In \cite{Gutperle:2001mb} it has been conjectured that the end point
of the decay of Type 0A/B due to the bulk tachyon in these theories
is the supersymmetric Type IIA/B theory. These arguments however are
dependent on the equivalance of certain M-theory backgrounds with
Type 0 and Type II in certain Melvin magnetic backgrounds. (We will
review these arguments later). Precisely at the point where one has
the flat 0A background however the region in which this equivalence
holds shrinks to zero. This renders the picture of magnetic flux shielding
considerably less trivial. On the other hand in a very recent paper
by David et al \cite{David:2001vm}, sigma model RG arguments have
been used to show that the end point of the decay of 0A in flat space
is IIA in flat space. How then can we reconcile the argument of the
previous paragraph with this claim?

The point is that sigma model arguments are made in a particular background
and give a set up in which perturbation theory around that background
can be done. One can calculate S-matrix elements for arbitrary numbers
of particles on the assumption that the coupling is so weak that the
back reaction on the background can be ignored. Of course if the coupling
were exactly zero there would be no back reaction and as the tachyon
slides down the potential there will be no change in the background
and it is consistent to argue that the end point is indeed IIA. However
in this paper we are interested in the question of what happens to
non-supersymmetric theories at finite (non-zero) coupling. In this
case one really needs to take into account the discussion of the previous
paragraph. In fact even if the tachyon potential bottomed out, at
any non-zero coupling the best that one could hope for is to end up
with a SUSY string theory (say IIA for 0A) in AdS space. 

What then might be the endpoint of this decay (for any theory with
bulk tachyons - not just 0A/B) at non-zero coupling? For now we note
that another endpoint seems plausible, that of space-time annihilation.
The motivation for this conjecture comes from a semi-classical argument
first presented by Witten \cite{Witten:1982gj}and directly parallels
the case for \( D \)-\( \overline{{D}} \) annihilation. This instability
dissappears in the zero coupling limit and so may only be associated
with a tachyonic instability at non-zero coupling.

A precise formulation of the conjecture is as follows: Suppose we
have a theory {\bf{A}} on a background {\bf{X}} which admits a semi-classical
instability (determined by an analysis of the low-energy effective
field theory). Now also assume that this semi-classical instability
varies smoothly as the moduli determining {\bf{X}} are varied. If
by adjusting the moduli of {\bf{X}} we reach a region in which the
semi-classical analysis is invalid (but otherwise would lead to an
instability), and we are instead afforded a perturbative description
of the quantum theory {\bf{Y}}, then the semi-classical instability
should be reflected by a tachyonic instability in the perturbative
description {\bf{Y}}. In addition, the endpoint of both instabilities
are to be identified.

We may consider a strong and weak form of the conjecture above distinguished
as follows:

Strong: A semiclassical instability as described above predicts the
existence of a tachyonic perturbative description of the resulting
theory and we should identify the endpoint of condensation of the
tachyon with that of the semiclassical instability.

Weak: A semiclassical instability as described above can be related
to the tachyonic mode whenever it exists in a perturbative description
of the resulting theory by identification of the endpoints of the
instabilities.

The weak form allows for theories which do not reduce to tachyonic
perturbative descriptions. We will see examples of each case below. 

The conjecture above will in many cases involve extrapolations from
strong coupling regions. As the systems are non-supersymmetric, such
extrapolations are unprotected and hence we do not know how to prove
them. Support for this conjecture comes from three directions. 1.
The close analogy to open string tachyons in unstable \( D \)-brane
systems for which there is much support. 2. The existence of closed
string tachyons in perturbative limits of systems exhibiting semi-classical
instabilities. 3. The fact that flat space decay endpoints do not
seem plausible for finite coupling as argued above. 

The outline of this paper is as follows. In the next section (\ref{sec:1110})
we will first apply the conjecture in systems with an eleven-dimensional
starting point. These are important because their perturbative limits
involve the well known ten-dimensional string theories. Starting with
semi-classically unstable circle and interval compactifications of
M-theory we identify the tachyonic perturbative limits involving Type
0A/B and nonsupersymmetric heterotic strings on flat backgrounds,
Melvin magnetic backgrounds, and noncompact orbifolds. We then move
on in (\ref{sec:109}) to similar considerations for ten-dimensional
starting points which admit a greater degree of control and in many
instances may be related to the eleven-dimensional cases by a {}``9-11{}''
flip duality. Along the way we will encounter several situations for
which recent analyses have led to conflicting conclusions and we will
discuss these issues. We end with some conclusions (in particular
for using the Scherk-Schwarz) mechanism for SUSY breaking and directions
for future work.

\section{Applications of the conjecture}

\subsection{11D \protect\( \to \protect \) 10D\label{sec:1110}}

In this section we will discuss circle and interval compactifications
of M-theory. The semi-classical instabilities arise in the eleven-dimensional
low-energy gravity theory as a result of the Kaluza-Klein structure
of the vacuum. We adapt several results from \cite{Costa:2000nw}
to the case of ten noncompact dimensions and discuss our own ideas
on the relevance of the semiclassical decay evolution. Identification
of perturbative string limits requires an extrapolation from strong
to weak coupling and in the absence of supersymmetry is unprotected.

\subsubsection{Twisted Circle \protect\( M^{10}\times S^{1}_{R,B}\protect \)\label{sec:TwistedCircle}}

Consider eleven-dimensional M-theory on a background which locally
resembles \( M^{10}\times S^{1} \)

\begin{equation}
\label{elevenmetric}
ds_{11}^{2}=-dt^{2}+d\rho ^{2}+\rho ^{2}d\phi ^{2}+dy_{i}dy^{i}+dx_{11}^{2}\qquad i=3,...,9
\end{equation}
 but differs globally by the nontrivial identifications: 

\begin{eqnarray}
x_{11} & \sim  & x_{11}+2\pi n_{1}R\label{twistedidentifications} \\
\phi  & \sim  & \phi +2\pi n_{1}BR+2\pi n_{2}\nonumber 
\end{eqnarray}
 We designate such a {}``twisted{}'' circle by \( S^{1}_{R,B} \).
Let us choose a periodic spin structure for the \( S^{1}_{R} \).
The twist parameter \( B \) takes values \( 0\leq |B|<\frac{2}{R} \).
For \( B=0 \) this is a supersymmetric compactification while for
\( B\neq 0 \) the spacetime supersymmetry is completely broken. The
effective theory governing the low-energy dynamics will generically
incorporate Einstein-Hilbert gravity. It has been known for some time
that gravity on a Kaluza-Klein background of this form exhibits a
semi-classical instability towards the annihilation of spacetime (first
discussed for five dimensions in \cite{Dowker:1995gb, Dowker:1996sg}
and later extended to eleven dimensions in \cite{Costa:2000nw}).
This instability is mediated by a bounce solution that takes the form
of an eleven-dimensional Euclidean Kerr black hole solution:\begin{eqnarray}
ds^{2}_{11}=(1-\frac{\mu }{r^{6}\Sigma })dx^{2}_{0}-\frac{2\mu \alpha sin^{2}\theta }{r^{6}\Sigma }dx_{0}d\phi +\frac{\Sigma }{r^{2}-\alpha ^{2}-\mu r^{-6}}dr^{2}+\Sigma d\theta ^{2} &  & \label{KerrBounce} \\
+\frac{sin^{2}\theta }{\Sigma }[(r^{2}-\alpha ^{2})\Sigma -\frac{\mu }{r^{6}}\alpha ^{2}sin^{2}\theta ]d\phi ^{2}+r^{2}cos^{2}\theta (d\chi ^{2}+sin^{2}\chi d\Omega ^{2}_{6}) &  & \nonumber 
\end{eqnarray}

where \( \Sigma \equiv r^{2}-\alpha ^{2}cos^{2}\theta  \), \( \mu  \)
is the black hole mass parameter, \( \alpha  \) is a single complexified
angular momentum parameter, and we have written \( d\Omega _{7} \)
as \( d\chi ^{2}+sin^{2}\chi d\Omega _{6} \) for later convenience.
The identifications (\ref{twistedidentifications}) are most easily
expressed in eleven-dimensional \( SO(2) \)-coordinates on \( S_{1}: \)\[
t,\rho ,\phi ^{(2\pi )},x_{3,}x_{4},x_{5},x_{6},x_{7},x_{8},x_{9},x_{11}\]
 but the instanton is more easily expressed in \( SO(9) \)-coordinates
on \( S_{8}: \) \[
x_{0},r,\theta ^{(\frac{\pi }{2})},\phi ^{(2\pi )},\chi ^{(2\pi )},\theta ^{(\pi )}_{1},\theta ^{(\pi )}_{2},\theta ^{(\pi )}_{3},\theta ^{(\pi )}_{4},\theta ^{(\pi )}_{5},\theta ^{(\pi )}_{6}.\]
 To match the bounce solution (\ref{KerrBounce}) to the unstable
background (\ref{elevenmetric},\ref{twistedidentifications}) the
background parameters must satisfy:

\begin{eqnarray}
R & = & \frac{\mu }{4r_{H}^{7}-3\alpha ^{2}r_{H}^{5}}\label{matching} \\
B & = & \frac{\alpha r_{H}^{6}}{\mu }-\frac{\alpha }{|\alpha |R}\nonumber 
\end{eqnarray}
 where \( r_{H} \) is the location of the Euclidean black hole horizon
satisfying:

\begin{equation}
\label{horizon}
r_{H}^{2}=\alpha ^{2}+\frac{\mu }{r_{H}^{6}}.
\end{equation}
 The coordinate singularity sets a lower bound on the range of the
radial coordinate\begin{equation}
\label{lowerr}
r\geq r_{H}.
\end{equation}
To estimate the decay rate we evaluate the Euclidean action \( I \)
for the bounce solution (\ref{KerrBounce}) and calculate \( \Gamma \sim e^{-I} \).
This evaluates to: \begin{equation}
\label{rate}
\Gamma \sim e^{-\frac{\pi ^{5}\mu R}{96G_{11}}}
\end{equation}
 where \( G_{11} \) is the eleven-dimensional Newton's constant.
Evaluating the decay rate in terms of the background parameters \( R,B \)
involves untangling expressions (\ref{matching},\ref{horizon}) which
can be quite difficult. The task simplifies for two important parameter
regions \cite{Dowker:1996sg}:

For \( |B|\sim 0 \) the expression for \( \mu  \) reduces to \( \mu =(\frac{7}{2})^{7}\frac{R}{|B|^{7}} \)
which clearly diverges for \( |B|\to 0. \) The decay rate vanishes
rendering the theory stable against the semi-classical instability.

For \( |B|=\frac{1}{R} \) (corresponding to \( \alpha =0 \) ) the
expression simplifies to \( \mu =(4R)^{8} \). One can demonstrate
that this is actually a minimum of \( \mu (B) \) and hence represents
the most unstable background.

In fact we have found with some numerical work that the decay rate
is a monotonically decreasing function of \( R \) for fixed \( B\neq 0 \)
indicating the expected stability of the decompactified theory and
in turn the maximum instability of the theory as \( R\to 0 \). For
fixed \( R \) one can also demonstrate that the decay rate is a monotonically
increasing function for \( B \) increasing from \( 0\to \frac{1}{R} \)
beyond which it monotonically decreases as \( B \) approaches \( \frac{2}{R}. \) 

The evolution of the background (\ref{elevenmetric},\ref{twistedidentifications})
after the decay is determined by finding a zero-momentum surface in
the bounce solution and using this as inital data for an analytic
continuation back to Lorentzian signature. Such a zero-momentum surface
is given by \( \chi =\frac{\pi }{2} \), so we may continue (\ref{KerrBounce})
by sending \( \chi \rightarrow \frac{\pi }{2}+i\tau  \) to obtain\begin{eqnarray}
ds_{11}^{2}=(1-\frac{\mu }{r^{6}\Sigma })dx_{0}^{2}-\frac{2\mu \alpha sin^{2}\theta }{r^{6}\Sigma }dx_{0}d\phi +\frac{\Sigma }{r^{2}-\alpha ^{2}-\mu r^{-6}}dr^{2}+\Sigma d\theta ^{2} &  & \label{ContinuedKerrBounce} \\
+\frac{sin^{2}\theta }{\Sigma }[(r^{2}-\alpha ^{2})\Sigma -\frac{\mu }{r^{6}}\alpha ^{2}sin^{2}\theta ]d\phi ^{2}+r^{2}cos^{2}\theta (-d\tau ^{2}+cosh^{2}\tau d\Omega _{6}) &  & \nonumber 
\end{eqnarray}

To get a feel for what the metric above describes let us first identify
the spatial infinity limit with the pre-decay geometry. This is nontrivial
owing to the double analytic continuation \( (t\rightarrow ix_{0},\chi \rightarrow \frac{\pi }{2}+i\tau ) \)
that we have used to get to this expression. Just after the decay
the geometry far from the decay nucleus should be in its pre-decay
form. Evaluating (\ref{ContinuedKerrBounce}) for \( r\rightarrow \infty  \)
we find\begin{equation}
\label{AsymptoticContiduedKerBounce}
ds^{2}_{11}(r\rightarrow \infty )\sim dx_{0}^{2}+dr^{2}+r^{2}d\theta ^{2}+r^{2}sin^{2}\theta d\phi ^{2}-r^{2}cos^{2}\theta d\tau ^{2}+r^{2}cos^{2}\theta cosh^{2}\tau d\Omega _{6}.
\end{equation}
 In this form it is not obvious that this metric describes asymptotically
flat space. To see this we first introduce radial coordinates \( (\widehat{\rho },\widehat{r}) \)
defined by\begin{eqnarray}
\widehat{\rho } & = & rsin\theta \label{SO2SO6coordinates} \\
\widehat{r} & = & rcos\theta \nonumber 
\end{eqnarray}
 and then introduce {}``flat{}'' coordinates

\begin{eqnarray}
\widetilde{r} & = & \widehat{r}cosh\tau \label{flatcoordinates} \\
\widetilde{\tau } & = & \widehat{r}sinh\tau .\nonumber 
\end{eqnarray}
In these coordinates (\ref{AsymptoticContiduedKerBounce}) takes the
form\begin{equation}
\label{finallyflat}
ds^{2}_{11}(r\rightarrow \infty )\sim dx_{0}^{2}+d\widehat{\rho }^{2}+\widehat{\rho }^{2}d\phi ^{2}+d\widetilde{r}^{2}+\widetilde{r}^{2}d\Omega _{6}-d\widetilde{\tau }^{2}
\end{equation}
which clearly describes a flat eleven-dimensional spacetime. 

The full post-decay metric expressed in these flat coordinates is
extremely complicated, however the most striking feature of the decay
scenario is easily seen as a result of (\ref{flatcoordinates}). Note
that the coordinate redefinitions imply

\begin{equation}
\label{hyperboliccondition}
\widetilde{r}^{2}+\widehat{\rho }^{2}-\widetilde{\tau }^{2}=r^{2}.
\end{equation}

A trivial algebraic rearrangement of (\ref{hyperboliccondition})
combined with the coordinate minimum for \( r \) in (\ref{lowerr})
implies the existence of a totally geodesic submanifold which is growing
in time

\begin{equation}
\label{bubble}
(\widetilde{r}^{2}+\widehat{\rho }^{2})_{min}=r^{2}_{H}+\widetilde{\tau }^{2}.
\end{equation}
 For the coordinate region inside of the expanding bubble \( \widetilde{r}^{2}+\widehat{\rho }^{2}<r^{2}_{H}+\widetilde{\tau }^{2} \)
the metric degrees of freedom cease to exist. This is the {}``bubble
of nothing{}'' annihilation of spacetime first described by Witten\cite{Witten:1982gj}.\footnote{%
Witten analyzed the spherically symmetric case corresponding to the
Euclidean-Scwarzchild bounce (the \( \alpha =0 \) form of (\ref{KerrBounce})).
In that case the expanding bubble has a spherically symmetric geometry.
Though the expanding surface in (\ref{bubble}) is in terms of a coordinate
sphere, the geometry of the expanding surface as measured by the metric
will be deformed from an \( SO(9) \) isometry to \( SO(2)\times SO(7) \).
} It is a difficult picture to consider but is strikingly reminiscent
of the idea of a purely topological phase of gravity. One should here
consider the corresponding story for unstable open string theories
in which the decay (either via condensation of tachyons or sphaleron
mediated semi-classical processes) often leads to an annihilation
of the open string degrees of freedom. For unstable D-branes one always
has the closed string vacuum to leave behind, but for an unstable
closed string vacuum the natural result seems, though perfectly analogous,
considerably more catastrophic. 

A great deal of discussion has been aimed at elucidating the picture
of this semi-classical decay in terms of a dimensionally reduced theory\cite{Dowker:1995gb, Dowker:1996sg, Gutperle:2001mb, Costa:2000nw}.
This has led to a number of seemingly strange equivalences. A very
simple example involves two different ten-dimensional descriptions
of the same eleven-dimensional process.\footnote{%
The example here is the shifted and unshifted reductions to ten-dimensions
of the critically twisted circle background first discussed in \cite{Dowker:1995gb}.
These reductions will be discussed in more detail in the next section.
} In one case the decay involves spacetime falling into a pointlike
singularity at an ever increasing rate, while the other description
resembles the (considerably less catastrophic) shielding of a Kaluza-Klein
magnetic field via pair production of magnetic monopoles.\footnote{%
We would like to thank Steve Giddings for discussion on this point.
} While these are certainly very interesting results, we take here
the view that exactly when a Kaluza-Klein reduction becomes appropriate
we lose the eleven-dimensional classical gravity approximation used
in these calculations. At sufficiently small length scales quantum
M-theory effects become important. The appropriate quantum description
in many cases will be in terms of a perturbative string theory on
the reduced background. We shouldn't concern ourselves with the dimensionally
reduced picture of the semi-classical instability. Instead we should
look for perturbative manifestations of this instability. Why then
identify the endpoints of the semi-classical and perturbative decays?
Again a chief motivation is the analogy with unstable open string
theories where we see the semiclassical instability described in \cite{Callan:1998kz, Savvidy:1998xx}
go over to the tachyonic instability elucidated by Sen in \cite{Sen:1998sm, Sen:1998ii, Sen:1998rg}. 

If the size of the radius shrinks below the eleven-dimensional Planck
length \( R<l_{P} \) then the eleven-dimensional gravity approximation
used above breaks down. However we are in most cases afforded a description
of the resulting dynamics in terms of weakly coupled string theory.
For \( B=0 \) this of course reduces to supersymmetric Type IIA strings
in a flat background. The bounce action above diverges for this particular
case reflecting that the eleven-dimensional theory is actually supersymmetric
and hence stable. We will now move on to the unstable \( B\neq 0 \)
cases and discuss their perturbative limits.

\subsubsection{Melvin Models\label{sec:MelvinModels}}

For values of \( 0\leq |B|<\frac{1}{R} \) and \( R<l_{P} \) we may
reduce the background (\ref{elevenmetric},\ref{twistedidentifications})
along the Killing vector \( l=\partial _{x_{11}}-B\partial _{\phi } \)
to obtain a Melvin magnetic flux tube background (a fluxbrane)\cite{Dowker:1995gb, Costa:2000nw}
which is described by:

\begin{equation}
\label{melvinmetric}
ds_{10}^{2}=\Lambda ^{\frac{1}{2}}(-dt^{2}+d\rho ^{2}+dy_{i}dy^{i})+\Lambda ^{-\frac{1}{2}}\rho ^{2}d\widetilde{\phi }^{2}
\end{equation}

\begin{equation}
\label{melvindilaton}
e^{\frac{4\Phi }{3}}=\Lambda =1+\rho ^{2}B^{2}
\end{equation}

\begin{equation}
\label{melvingaugefield}
A_{\widetilde{\phi }}=\frac{B\rho ^{2}}{2\Lambda }\quad \Longrightarrow \quad \frac{1}{2}F_{\mu \nu }F^{\mu \nu }|_{\rho =0}=B^{2}
\end{equation}
where \( \widetilde{\phi }\equiv \phi -Bx_{11} \). This curved ten-dimensional
background incorporates an axially symmetric RR two-form field strength
parameterized by its central (\( \rho =0 \)) value \( B \), and
a nontrivial dilaton which grows as we move away from the \( \rho =0 \)
hyperplane for \( B\neq 0. \) To determine the perturbative content
of the theory we should recall that the eleven-dimensional starting
point was M-theory on a flat Kaluza-Klein background. For the periodic
choice of spin structure on the \( S^{1}_{R} \) factor and for \( B=0 \)
this reduces to Type IIA strings on \( M^{10} \) as discussed above.
For \( B\neq 0 \) we should then obtain Type IIA strings propagating
on the Melvin background.\footnote{%
By Type IIA strings we mean a theory of closed oriented world sheets
with the standard Type IIA GSO constraints quantized in this background.
}

Quantizing strings on the background (\ref{melvinmetric},\ref{melvindilaton},\ref{melvingaugefield})
faces the twin difficulties of incorporating RR flux and a curved
geometry and is beyond current understanding. Applying the strong
form of the conjecture discussed in the introduction would however
imply that the corresponding closed string fluctations should admit
at least one tachyonic mode whose condensation would also lead to
the annihilation of the spacetime. 

\begin{figure}[!h]
{\centering \includegraphics{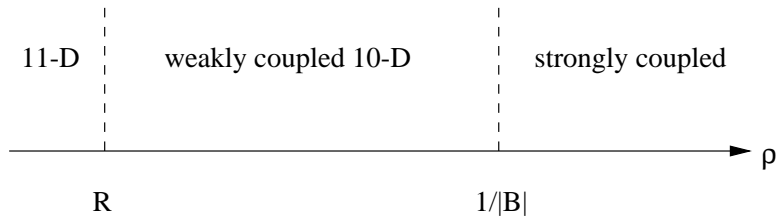} \par}

\caption{Radial range of weakly coupled 10D physics for a noncritical Melvin
background. \label{fig:VALID}}
\end{figure}

The theory for \( |B|\neq 0 \) is continuously connected to the supersymmetric
Type IIA vacuum. It may seem natural that condensation of a closed
string tachyon would in this case relax the value of \( |B| \) to
zero, restoring the supersymmetric vacuum.\footnote{%
This is the identification made in \cite{Gutperle:2001mb}
} In this sense the Melvin magnetic flux would represent an excited
state in the Type IIA theory, decaying by flux dissipation.\footnote{%
This has been the motivation behind several assertions that the tachyonic
Melvin background decays into the {}``underlying{}'' supersymmetric
vacuum.
} However the Melvin background does not merely constitute weakly coupled
Type IIA string theory with some additional unstable flux. As one
can see from the nontrivial dilaton profile (\ref{melvindilaton})
a description in terms of any weakly coupled string theory will only
be possible in the spatial region \( R<\rho <\frac{1}{|B|}, \) see
figure \ref{fig:VALID}.

For \( \rho <R \) we invalidate the Kaluza-Klein ansatz, while for
\( \rho >\frac{1}{|B|} \) the string coupling becomes strong. In
either case we must utilize the eleven-dimensional description.

\subsubsection{Critical Melvin and Type 0A:The Scherk-Schwarz Circle\label{sec:CriticalMelvin}}

For \( |B|=\frac{1}{R} \)the effect of the twisted identifications
(\ref{twistedidentifications}) is to accompany a \( 2\pi R \) translation
in \( x_{11} \) (generated by \( n_{1}\to n_{1}+1) \) with a \( 2\pi  \)
rotation in \( \phi  \). This forces fermions to pick up a \( -1 \)
when transported around the compact circle (so called Scherk-Schwarz
(SS) boundary conditions \cite{Scherk:1979ta}) and leaves bosons
unaffected. Our starting point was a periodic spin structure on the
\( S^{1}_{R} \) so the net effect of \( |B|=\frac{1}{R} \) is to
exactly reverse this choice of spin structure. Thus one may consider
the {}``critical{}'' case of \( |B|=\frac{1}{R} \)in either of
two ways:

a. Periodic spin structure on \( S^{1}_{R} \) and \( |B|=\frac{1}{R}. \) 

\begin{figure}[!h]
{\centering \includegraphics{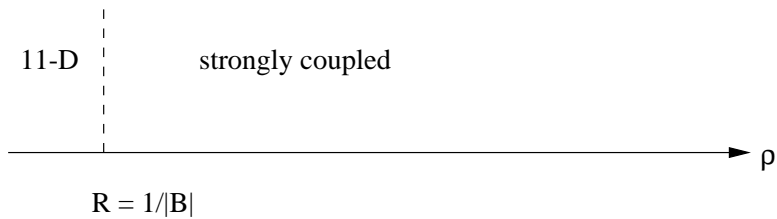} \par}

\caption{Radial range of perturbative descriptions for a critical Melvin background.\label{fig:CRITVALID}}
\end{figure}

This case will again reduce to Type IIA strings propagating on a Melvin
magnetic background. However in this critical case (figure \ref{fig:CRITVALID})
the theory is nowhere described by a weakly coupled ten-dimensional
string theory since the relevant region shrinks to zero.

b. Antiperiodic spin structure on \( S^{1}_{R} \) and \( |B|=0 \).

In this case the resulting ten-dimensional background is flat \( M^{10} \).
Bergman and Gaberdiel have considered M-theory compactified on a Scherk-Schwarz
circle and conjectured that the appropriate perturbative degrees of
freedom are Type 0A strings\cite{Bergman:1999km}.\footnote{%
This relies on interpolating orbifolds of the form \( \Sigma _{R}=\frac{S^{1}_{R}}{(-1)^{F_{s}}\times S} \)
where \( F_{s} \) is the spacetime fermion number and \( S \) generates
a half shift around the \( S^{1}_{R} \). This amounts to compactifying
on a circle of radius \( R \) with Scherk-Schwarz boundary conditions.
The key to this construction is that in the limit \( R\to 0 \) the
orbifold factorizes into a compactification on a circle of radius
\( 2R \) and a twist by \( (-1)^{F_{s}} \) . The compactification
gives IIA on \( S^{1}_{2R} \) and then applying \( (-1)^{F_{s}} \)
twists this into 0A on \( S^{1}_{2R.} \) The factor of \( 2 \) arises
from the half shift and becomes irrelevant in the decompactification
limit. An extensive application of these interpolating models for
nonsupersymmetric string theories is in \cite{Blum:1997cs}\cite{Blum:1998gw}.
} While this conclusion is still unverified it agrees with our conjecture
in the sense that the spectrum of Type 0A strings on \( M^{10} \)
admits a closed string tachyon. The endpoint of condensation of the
Type OA closed string tachyon would then be identified with the annihilation
of spacetime. 

Thus far only the special cases \( |B|=0,\frac{1}{R} \) admit realiable
perturbative information. The presence of RR flux and the curvature
of spacetime for \( |B|\neq 0,\frac{1}{R} \) renders even a spectrum
calculation beyond our reach, however, we will later discuss similar
models in nine dimensions for which the full spectrum is trivially
obtained.

\subsubsection{Noncompact Orbifolds\label{sec:NoncompactOrbifold}}

When the twist parameter takes special values of the form \( |B|=\frac{1}{NR} \)
or \( |B|=\frac{1}{R}+\frac{1}{NR} \) we can perform \( SL(2,Z) \)
transformations on the (\( x_{11},\phi  \)) compactification torus
and reduce to a ten-dimensional background of the form\cite{Gutperle:2001mb}:

\begin{equation}
\label{orbifoldmetric}
ds_{10}^{2}=\eta _{\mu \nu }dx^{\mu }dx^{\nu }+d\widehat{r}^{2}+\frac{\widehat{r}^{2}}{N^{2}}d\widehat{\phi }^{2}
\end{equation}

\begin{equation}
\label{orbifolddilaton}
e^{\frac{4\Phi }{3}}=N^{2}
\end{equation}

\begin{equation}
\label{orbifoldgaugefield}
A_{\widehat{\phi }}=\frac{1}{2}R\frac{N-1}{N}
\end{equation}
 The resulting spacetime is that of a noncompact orbifold with the
fundamental region a cone of deficit angle \( \frac{2\pi }{N} \).
The dilaton in this case is constant throughout the spacetime. If
we start with periodic spin structure on the \( S^{1}_{R} \) then
for \( |B|=\frac{1}{NR} \) the correct perturbative degrees of freedom
involve Type 0A strings. For periodic spin structure on the \( S^{1}_{R} \)
and \( |B|=\frac{1}{R}+\frac{1}{NR} \) we transform to \( B'=B-\frac{B}{|B|}\frac{1}{R} \)
and reverse the spin structure as in section \ref{sec:CriticalMelvin}.
This reduces to Type IIA strings propagating on the orbifold background
(\ref{orbifoldmetric},\ref{orbifolddilaton},\ref{orbifoldgaugefield}).

Even though the geometry is locally flat, string quantization on this
background is difficult owing to the presence of the RR Wilson line.
Our conjecture would imply the existence of a tachyonic instability
in the perturbative description for either \( |B|=\frac{1}{NR} \)or
\( |B|=\frac{1}{R}+\frac{1}{NR} \) with the endpoint of its condensation
involving the annihilation of spacetime.

This result seems to contradict the conclusions reached in \cite{Adams:2001sv}
where it was found that the effect of tachyon condensation in noncompact
orbifolds is to {}``un-orbifold{}'' the theory restoring the {}``underlying{}''
supersymmetric closed string vacuum. This deserves some discussion.
Among the arguments in \cite{Adams:2001sv} was the observation that
the orbifold fixed plane represents a curvature singularity in a locally
flat spacetime which may be viewed as a localized excitation above
the underlying background. In fact for the special cases \( Z_{N} \)
with \( N \) odd it was pointed out that the closed string tachyons
are localized to the orbifold fixed plane and by an analogy with open
string tachyons localized to unstable \( D \)-branes represent an
instability towards decay of the localized energy density restoring
the supersymmetric vacuum. The elegant analysis in \cite{Adams:2001sv}
is sound, however when one tries to apply their conclusions to the
present scenario we find some obstacles.

First of all the analysis was performed in the zero coupling limit.
As we argued in the introduction, for zero coupling the tachyon may
condense without affecting the underlying background. We expect that
for non-zero coupling the tachyon will have a more dramatic effect
on the background. Secondly to connect these orbifolds to the twisted
circle compactifications discussed above one must include the RR Wilson
line (\ref{orbifoldgaugefield}). In \cite{Adams:2001sv} the spectrum
of the theory was computed without incorporating any Wilson line.
As we have mentioned earlier including the RR Wilson line is difficult,
however one can circumvent this difficulty by looking at the corresponding
situation obtained by compactifying a ten-dimensional theory on a
twisted circle. In this case the Wilson line will arise in the NSNS
sector and the spectrum can be evaluated exactly. We will discuss
these in more detail in section \ref{sec:109} but for now we point
out that one important effect of including the Wilson line is the
localization of closed string tachyons for any value of \( 0<|B|<\frac{1}{R} \)
(particularly \( |B|=\frac{1}{NR} \) with \( N \) even or odd).
In addition the {}``curvature singularity as a localized excitation
above a flat background{}'' argument needs to be reconsidered. For
a Wilson line of the form (\ref{orbifoldgaugefield}) the ten-dimensional
geometry actually lifts to a flat eleven-dimensional geometry which
is everywhere regular. Probing distances very close to the orbifold
fixed plane invalidates the Kaluza-Klein ansatz and we should replace
the reduced theory by its eleven-dimensional interpretation.

\subsubsection{The Scherk-Schwarz Interval \protect\( M^{10}\times I^{SS}_{L}\protect \)
\label{sec:Scherk-SchwarzInterval}}

Consider now Horava-Witten (HW) theory \cite{Horava:1996qa, Horava:1996ma},
i.e. eleven-dimensional M-theory compactified on a line segment of
length \( L \). In addition to the bulk degrees of freedom anomaly
cancellation requires an \( E8 \) gauge theory to live on each ten-dimensional
wall bounding the bulk spacetime. Fabinger and Horava (FH) considered
the scenario that results from reversing the chirality of fermions
living on one of the walls\cite{Fabinger:2000jd}. This breaks the
spacetime supersymmetry and renders the theory unstable. FH demonstrated
the existence of an attractive casimir force between the walls and
then went on to discuss a semi-classical instability towards formation
of a wormhole-like tube connecting the two walls, the interior of
which has no metric degrees of freedom. This tube grows radially outward
eating up both the \( E8 \) walls and the bulk spacetime. This system
is equivalent to a compactification of M-theory on a Scherk-Schwarz
circle of radius \( \frac{L}{\pi } \) followed by a \( Z_{2} \)
orbifolding. This may be viewed as a \( Z_{2} \) orbifolding of the
critically twisted circle \( S^{1}_{\frac{L}{\pi },\frac{\pi }{L}} \)
discussed in section \ref{sec:CriticalMelvin}. The relevant bounce
solution is simply the \( Z_{2} \) invariant form of (\ref{KerrBounce})
evaluated at \( R_{FH}=\frac{L}{\pi } \) and \( B=\frac{\pi }{L} \).

\begin{equation}
\label{IntervalBounceMetric}
ds_{11}^{2}=(1-\frac{\mu }{R^{8}})dx_{11}^{2}+(1-\frac{\mu }{R^{8}})^{-1}dr^{2}+r^{2}(d\chi ^{2}+sin^{2}\chi d\Omega _{8})
\end{equation}
For the critical case we may easily express the mass parameter \( \mu  \)
in terms of the background parameters\begin{equation}
\label{intervalmu}
\mu =(\frac{4L}{\pi })^{8}.
\end{equation}
 Borrowing expression (\ref{rate}) for the decay rate we find\begin{equation}
\label{intervalrate}
\Gamma \sim e^{-\frac{2^{11}L^{9}}{3\pi ^{4}G_{11}}}.
\end{equation}
 Analysis of the post decay evolution proceeds along the lines of
section \ref{sec:TwistedCircle}. The picture is that of a \( Z_{2} \)
projection of Witten's spherically symmetric bubble of nothing expanding
in time. We now discuss two possible perturbative limits of the FH
scenario.

\subsubsection{The Case of the Shrinking Interval \protect\( M^{10}\times I^{SS}_{L\to 0}\protect \)
\label{subsec:ShrinkInterval}}

Consider the situation where the two \( E8 \) walls come together.
For \( L\sim l_{P} \) the eleven-dimensional gravity approximation
breaks down. We might anticipate a result similar to the HW case for
which the appropriate description as the two \( E8 \) walls come
together is in terms of weakly coupled Heterotic \( E8\times E8 \)
string theory (\( H^{susy}_{E8\times E8}) \) on \( M^{10} \). For
the case of FH the resulting perturbative string description must
have broken supersymmetry. Furthermore, our conjecture in its strong
form would imply that the resulting string theory should have a tachyonic
mode which mediates the annihilation of spacetime. There are seven
candidate nonsupersymmetric heterotic string theories \cite{Kawai:1986vd}.
Their relevant properties are summarized in the table below.

\vspace{0.3cm}
{\centering \begin{tabular}{|c|c|c|}
\hline 
Gauge Symmetry&
Tachyon Representation&
Chiral\\
\hline
\hline 
\( H_{SO(16)\times SO(16)} \)&
tachyon-free&
yes\\
\hline 
\( H_{SO(32)} \) &
\( (32_{v)} \)&
no\\
\hline 
\( H_{SO(8)\times SO(24)} \) &
\( (8_{v},1) \)&
yes\\
\hline 
\( H_{SU(16)\times SO(2)} \) &
\( (1,2_{v}) \)&
yes\\
\hline 
\( H_{SO(16)\times E8} \)&
\( (16_{v},1) \)&
yes\\
\hline 
\( H_{E7\times SU(2)\times E7\times SU(2)} \)&
\( (1,2,1,2) \)&
yes\\
\hline 
\( H_{E8} \) &
\( (1) \)&
no\\
\hline
\end{tabular}\par}
\vspace{0.3cm}

To identify the best candidate theory we consider the membrane world-volume
anomaly analysis of HW\cite{Horava:1996ma}. For a topologically stabilized
membrane wrapping a large \( S_{x^{9}}^{1} \) and stretched between
the two walls, the right-moving \( 8'' \) fermions\footnote{%
Right-moving here is simply \( - \) chirality under the \( SO(1,1) \)
isometry of the \( (t,x^{9}) \) cylinder, and \( 8'' \) is the conjugate
spinor of \( SO(8) \).
} induce a three-dimensional gravitational anomaly since the world-volume
has orbifold singularities, i.e. it is not a smooth manifold. To cancel
this anomaly one must add left-moving current algebra modes with \( c=16 \).
Since the anomaly is localized and evenly distributed between the
two boundaries of the world-volume, the current algebra modes should
be evenly distributed between the two ends as well. In the HW case
spacetime supersymmetry is preserved, and the only supersymmetric
string theory with this world-sheet structure is the \( H^{susy}_{E8\times E8} \)
theory.\footnote{%
More precisely, the \( H^{susy}_{E8\times E8} \) makes use of the
left-moving current algebra in two independent sets,e.g. in the free
fermionic construction the current algebra is carried by two independent
sets of 16 left moving fermions. The \( H^{susy}_{SO(32)} \) on the
other hand makes use of a single set of 32 left moving fermions. 
} If the spacetime supersymmetry is broken, as in the case at hand,
then we should look for nonsupersymmetric strings with this world-sheet
structure. Only two of the seven cases above are of this type; the
\( H_{E8\times SO(16)} \) and \( H_{SO(16)\times SO(16)} \) theories.
There are two additional reasons which lead to the choice of \( H_{E8\times SO(16)} \)
as the \( L\to 0 \) limit of FH. Motivated by our conjecture, we
choose the only one of these two that is tachyonic. This would follow
from the strong form of the conjecture. An indication that this is
plausible was worked out in \cite{Fabinger:2000jd} where the mass
of a membrane state stretched between the two walls was calculated
and shown to become tachyonic when the two walls are sufficiently
close, i.e. \( L<l_{P} \). Of course the membrane energy calculation
becomes invalid precisely in this regime, but it does seem indicative
of a continued instability of the theory. Furthermore, the GSO constraints
which lead to the \( H_{E8\times SO(16)} \) theory differ from those
that lead to the \( H^{susy}_{E8\times E8} \) theory by a twist of
\( exp(i\pi F_{B}) \) which affects only half of the left moving
current algebra modes.\footnote{%
In the free-fermionic construction the world-sheet fields are divided
into 8 left and 8 right-moving bosons \( (\partial X^{\mu },\widetilde{\partial X^{\mu })} \),
8 right-moving fermions \( \widetilde{\psi ^{\mu }} \), and two sets
of 16 left-moving fermions \( (\psi _{A},\psi _{B}) \). The \( H^{susy}_{E8\times E8} \)
GSO constraints involve \( exp(i\pi F_{A})=exp(i\pi F_{B})=exp(i\pi \widetilde{F})=+1 \),
whereas the \( H_{E8\times SO(16)} \) construction uses \( exp(i\pi F_{A})=exp(i\pi F_{B}+i\pi \widetilde{F})=+1 \),
where \( F_{A},F_{B},\widetilde{F} \) are the world-sheet fermion
numbers for \( \psi _{A},\psi _{B},\widetilde{\psi ^{\mu }} \) respectively.
} It may seem odd that the gauge group is broken asymmetrically as
the two-walls come together since there seems to be no {}``preferred{}''
wall. However both \( E8 \) and \( SO(16) \) require 16 current
algebra fermions on the corresponding worldsheet so in a sense the
walls are on equal footing. The effect of {}``flipping{}'' one of
the \( E8 \) walls on the M-theory side must translate into a modification
of the GSO projection on one of the two sets of 16 current algebra
fermions on the heterotic string worldsheet. In any case, the two
walls coming together ventures through intermediate coupling regions
(for which there is no known description) unprotected by supersymmetry,
rendering the specific mechanism behind the spacetime gauge symmetry
breaking difficult to study.

\subsubsection{The Case of the Shrinking Transverse Circle \protect\( M^{9}\times S^{1}_{R\to 0}\times I^{SS}_{L=finite}\protect \)\label{subsec:shrinkingtransverseinterval}}

\begin{figure}[!h]
{\centering \resizebox*{6in}{!}{\includegraphics{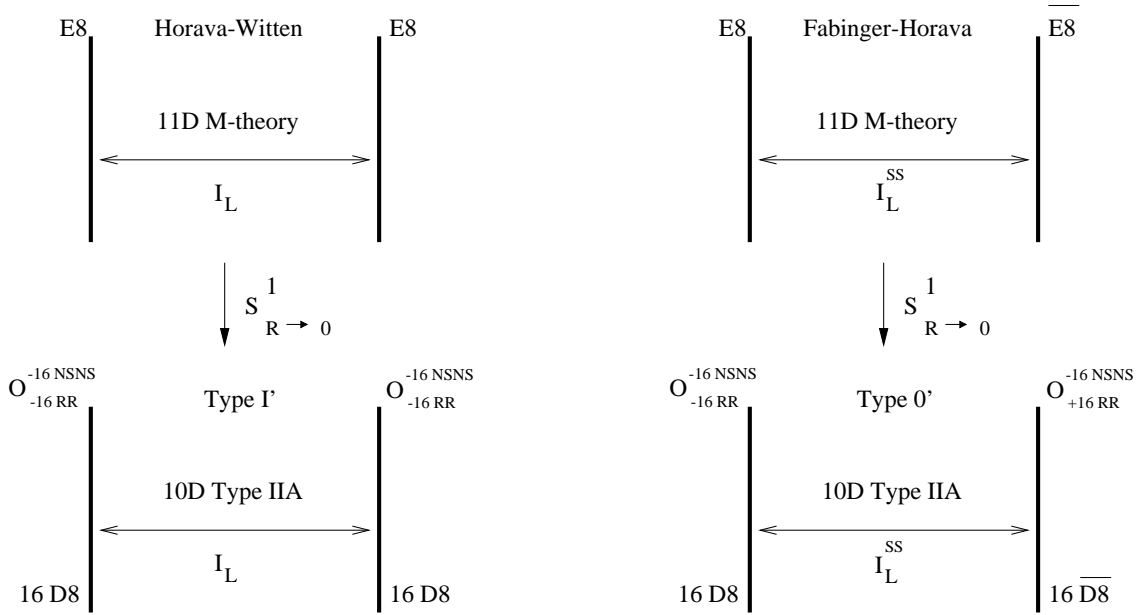}} \par}

\caption{Compactifications of HW and FH on transverse circles. The charges
carried by the orientifold planes are explicitly designated.\label{fig:HWFH}}
\end{figure}

Now consider the FH background keeping the Scherk-Schwarz interval
length \( L \) large and further compactifying the theory on a transverse
circle \( S^{1}_{R} \) with a periodic choice of spin structure.
For \( R<l_{P} \) we should be able to describe the system by a nonsupersymmetric
variant of the familiar Type I' theory (Type 0') as shown in figure
\ref{fig:HWFH}.

In the familiar supersymmetric case Type I' is obtained from IIA compactified
on a circle by dividing out by the \( Z_{2} \) symmetry \( g=I\Omega  \)
where \( \Omega  \) is the world sheet orientation reversal and \( I:\, x^{9}\rightarrow -x^{9} \).
In the original M-theory picture of the FH construction the reversal
of orientation of one of the \( \mathrm{E}_{8} \) walls may be accomplished
by dividing the theory by an additional \( Z_{2} \) symmetry generated
by \( g'=S(-1)^{F_{s}} \) giving\begin{equation}
\label{twoz2s}
M^{9}\times S^{1}_{R}\times S^{1}_{\frac{L}{\pi }}/(Z_{2}\times Z_{2}')
\end{equation}
where in the M-theory case the first \( Z_{2} \) is generated by
just the reflection. Thus the theory that we get on reduction to ten
dimensions \( R\to 0 \) should be Type IIA on \( \textrm{M}^{9}\times S^{1}_{\frac{L}{\pi }}/(Z_{2}\times Z'_{2}) \)
where the two symmetries are generated by \( g \) and \( g'. \)
This is equivalent to a theory defined in \cite{Antoniadis:1998ki, Kachru:1999ed}
and has an orientifold 8-plane at one fixed point and an anti-orientifold
at the other fixed point. There is no need to add \( D \)-branes
to cancel RR tadpoles but getting flat space would require the cancellation
of the NSNS tadpoles and thus would entail the presence of 16 \( D8 \)-branes
and 16 \( \overline{D8} \)-branes with the passage to the corresponding
M theory case possible when the former are coincident with the orientifold
plane and the latter with the anti-orientifold plane. Thus this limit
of the FH theory is simply an orientation reversed version of the
Type I' theory. We should then really be considering an \( S^{1} \)
compactification of FH with a Wilson line \( Y \) which breaks the
\( E8\times E8 \) gauge symmetry to \( SO(16)\times SO(16) \). This
has a potentially tachyonic mode coming from the twisted sector \cite{Bergman:1999km},
with mass\begin{equation}
\label{orbifoldtachyonmass}
m^{2}=\frac{L^{2}}{4\pi ^{2}\alpha '^{2}}-\frac{2}{\alpha '}
\end{equation}
 which becomes tachyonic when \( L<2\pi \sqrt{2\alpha '} \) . This
state clearly survives the \( g \) and \( g' \) projections.

It should be stressed that this is a closed string tachyon coming
from the lowest winding mode of the twisted sector of the theory.
In addition of course the theory has open string tachyons coming from
the open strings stretched between the \( D \)-\( \bar{D} \) when
they get within a distance \( \pi \sqrt{2\alpha '} \) of each other.
The \( D \)-branes are attracted to each other and will annihilate
due to this, leaving us with a background that will have a negative
cosmological constant (due to the negative tension of the orientifold
anti-orientifold system). What would one expect to be the end point
of the decay of the closed string tachyon?

The answer to this according to our conjecture should be obtained
from the picture of semi-classical vacuum bubble decay that one one
has when \( L>2\pi \sqrt{2\alpha '} \) in analogy with the corresponding
M-theory FH case. Thus as in the FH case one expects this theory to
be subject to space time annihilation. One might then conjecture that
for \( L<2\pi \sqrt{2\alpha '} \) also the end point of the tachyonic
decay should also be interpreted as the catastrophic annihilation
of the background, even though strictly speaking this is not a region
where the geometrical argument of Witten is directly applicable. 

In the region for which the lowest mode becomes tachyonic \( L<2\pi \sqrt{2\alpha '} \)
the nine-dimensional theory should be replaced by an appropriate T-dual
description. In the supersymmetric case the appropriate T-dual description
is the Type I theory. This theory can also constructed from Type IIB
by gauging the world-sheet parity \( \Omega  \) (Type IIB/\( \Omega  \))
and adding 32 \( D9 \)-branes to cancel the resulting massless RR
and NSNS tadpoles. At strong coupling this theory is described by
the weakly coupled \( H^{susy}_{SO(32)} \) theory. See figure \ref{fig:web1}
in the appendix.

For the T-dual of Type 0' one expects a perturbative description in
terms of a nonsupersymmetric analog of Type I theory. Before discussing
this theory in detail we use the strong form of our conjecture to
anticipate some of its features. The semi-classical instability in
eleven-dimensions annihilates both the gauge degrees of freedom and
the spacetime. In the compactification at hand the gauge degrees of
freedom and the spacetime are described by two different sectors of
the theory (the former by open strings and the latter by closed strings).
The single semi-classical annihilation instability of the eleven-dimensional
theory should then descend to two tachyonic instabilities, one leading
to the annihilation of the gauge degrees of freedom and the other
leading to the annihilation of spacetime.\footnote{%
This is in contrast to the discussion in section \ref{subsec:ShrinkInterval}.
For heterotic strings it is difficult to imagine an annihilation of
the gauge degrees of freedom (now carried by closed strings) without
losing all closed string degrees of freedom.
} 

To construct the T-dual theory we permute the two \( Z_{2} \) symmetries.
In the \( strict \) \( L\to 0 \) limit one is then left with Type
0B/\( \Omega  \), which has been called Type 0 theory (a nonsupersymmetric
analog of Type I). This theory has been constructed in \cite{Bianchi:1990yu, Sagnotti:1995ga, Bergman:1997rf, Bergman:1999km}
and indeed contains both open string tachyons charged under the gauge
symmetry as well as a closed string tachyon (the Type 0B tachyon survives
the \( \Omega  \) projection). The massless NSNS tadpole contribution
for Type 0B/\( \Omega  \) is twice that in Type IIB/\( \Omega  \),
so to formulate the Type 0B/\( \Omega  \) theory in flat space (which
we expect for the T-dual description of Type 0' in flat space) requires
the addition of 64 \( D9 \)-branes. The absence of massless RR tadpoles
implies that 32 of these should be \( D9 \)-branes and the other
32 \( \overline{D9} \)-branes. However the Type 0 theories exhibit
two types of \( Dp \)-brane for any given \( p \) distinguished
by their charge under the twisted sector fields \cite{Bergman:1997rf, Bergman:1999km}.
If we designate these by \( Dp' \) and \( Dp'' \), then the massless
tadpoles can be cancelled by adding \( n \) \( D9'-\overline{D9'} \)
pairs and \( 32-n \) \( D9''-\overline{D9''} \) pairs. The resulting
gauge symmetry is then \( SO(n)\times SO(n)\times SO(32-n)\times SO(32-n). \)
Of course we originally had the 16 \( D8 \)-branes and 16 \( \overline{D8} \)-branes
of Type 0', but these are standard Type II \( D8 \)-branes. For finite
radius on the T-dual side the situation can be described by the splitting
of Type II \( Dp \)-branes into pairs of Type 0 \( Dp'/Dp'' \)-branes
on the dual circle \cite{Imamura:1999um}, however some subtleties
for the strict \( L\to 0 \) limit are unresolved.\footnote{%
We thank Oren Bergman for several useful discussions on this and related
issues. 
} This is presently under investigation \cite{Bergman:Inprogress}.

This gauge symmetry enhancement is an unusual feature of Scherk-Schwarz
compactifications. The point is that for a Scherk-Schwarz compactification
there are new twisted sector states in the theory which only become
light as \( R\to 0 \). In the closed string sector these states form
an essential part of the conjectured relationship between Type 0A
and M-theory \cite{Bergman:1999km} where they become the additional
NSNS and RR fields of Type 0A (relative to Type IIA). In the case
of open strings they lead to the gauge symmetry enhancement discussed
above \cite{Bergman:Inprogress}.

\subsubsection{Type 0 at Strong Coupling}

\begin{figure}[!h]
{\centering \includegraphics{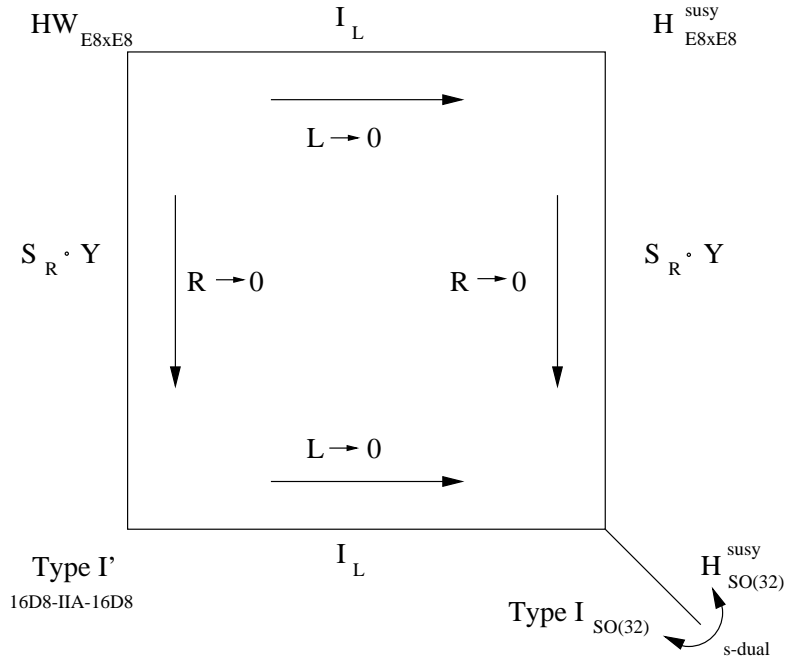} \par}

\caption{The standard web of dualities obtained by supersymmetric compactifications
of the Horava-Witten theory \cite{Horava:1996qa, Polchinski:1996df}.\label{fig:web1}}
\end{figure}

\begin{figure}[!h]
{\centering \includegraphics{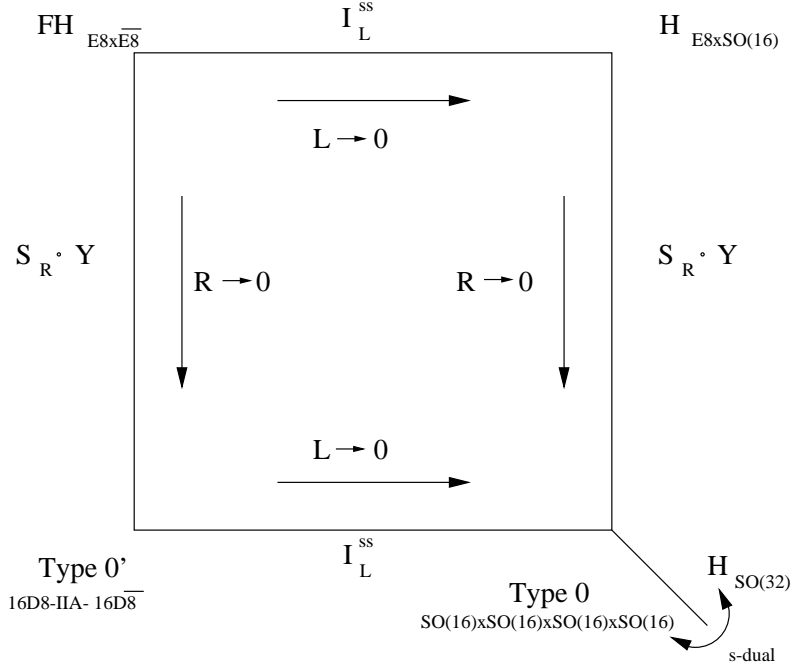} \par}

\caption{The conjectured web of dualities obtained from simple compactifications
of the Fabinger-Horava background, c.f. sections \ref{subsec:ShrinkInterval},\ref{subsec:shrinkingtransverseinterval}.\label{fig:web3}}
\end{figure}

Our analysis thus far has been based on perturbative constructions
and nonperturbative relations motivated by the tachyon/semi-classical
instability conjecture. At this point we will take an aside from the
main line of this paper to complete the picture that seems to emerge.
If we are bold enough to push the admittedly speculative results of
section \ref{subsec:shrinkingtransverseinterval} to strong coupling
we may expect that the S-dual of the Type 0 theory discussed above
will involve a nonsupersymmetric heterotic string theory with a tachyonic
instability which is charged under the gauge symmetry. The task then
is to identify which of the seven candidate theories is appropriate.
We should point out the similarity between the limits of FH that we
have been considering and the standard picture for the HW background
\ref{fig:web1}. Where the Type 0' theory resulted from compactifying
the FH theory on a circle with a Wilson line \( Y \) breaking the
\( E8\times E8 \) to \( SO(16)\times SO(16) \), we can consider
also compactifying the \( H_{E8\times SO(16)} \) theory on a circle
with the same Wilson line \( Y \). The resulting T-dual description
is the \( H_{SO(32)} \) theory \cite{Ginsparg:1987wr}. This leads
us to conjecture that the S-dual of Type 0 is described by the \( H_{SO(32)} \)
string as in figure \ref{fig:web3}.

An unusual feature of this S-duality proposal is the change in rank
of the gauge symmetry group. Though S-dualities of this type are know
in field theory \cite{Seiberg:1995pq}, we know of no such example
in string theory. Of course the S-duality relationship that we have
proposed has its geometric origin in a compactification torus which
involves a Scherk-Schwarz cycle. The subtleties involved in a strict
zero-radius limit for a Scherk-Schwarz circle should manifest itself
when taking the strong coupling limit of Type 0. This mismatch in
gauge group rank was pointed out in a closely related context in \cite{Blum:1997cs, Blum:1998gw}.
It also motivated the authors of \cite{Bergman:1997rf} to conjecture
that the strong coupling dual of the ten-dimensional Type 0 theory
is the \( D=26 \) bosonic string compactified on the \( SO(32) \)
lattice since this is the only possible closed string theory with
a rank 32 gauge group. Our conjecture stems from a larger scheme of
dualities (presented in figure \ref{fig:web3} in the appendix) akin
to the familiar web of dualities shown in figure \ref{fig:web1}.
We find these similarities very compelling and are presently working
to understand the gauge symmetry enhancement issue in more detail
\cite{Bergman:Inprogress}.

A standard technique for supporting S-duality conjectures is to find
a stable soliton that becomes light in the strong coupling limit and
identify its fluctuation spectra with that of the fundamental degrees
of freedom in the dual theory. In Type I/Heterotic \( SO(32) \) duality
for instance the massless fluctuations of the Type I \( D \)-string
(with mass inversely proportional to the string coupling) are identified
with world-sheet fields of the \( F \)-string in \( H^{susy}_{SO(32)} \)
\cite{Polchinski:1996df}. In particular the \( DD \) open string
modes become the \( F \)-string fields with spacetime quantum numbers
while the \( DN \) open string modes go over to the current algebra
degrees of freedom. 

Trying to apply this reasoning to the present case immediately confronts
an ambiguity in that there are two types of non-tachyonic \( D \)-string
present in Type 0 \cite{Bergman:1997rf, Bergman:1999km}. Furthermore
the fluctuation spectrum on either \( D \)-string does not match
up with the worldsheet structure of the \( H_{SO(32)} \) \( F \)-string.
The resolution of this ambiguity has already been suggested in \cite{Bergman:1999km}
based on observations noted earlier in \cite{Klebanov:1998yy}. The
appropriate soliton to consider is a bound state of the two \( D \)-strings
present in the theory. In particular the modes of open strings stretched
between the two \( D \)-strings give rise to the worldsheet fermions
carrying a spacetime vector index in the dual theory. These bound
states are very interesting on their own in so far as they are very
BPS-like despite being non-supersymmetric. For example when parallel
two of these bound states exhibit no force on one another in a manner
analogous to BPS \( D \)-strings. In addition the bound state is
decoupled from all of the twisted sector fields in the theory including
the bulk tachyon \cite{Klebanov:1998yy}. Similar proposals have been
discussed for the self-duality of Type 0B \cite{Bergman:1999km, Craps:2000zr}.
A thorough understanding of these soliton bound states would provide
considerable support for the picture that we have outlined and is
presently under investigation.

\subsection{10D \protect\( \to \protect \) 9D\label{sec:109}}

We now turn to applications of the conjecture for compactifications
from ten to nine dimensions. Some advantages over the previous discussion
are that any Kaluza-Klein gauge field will now reside in the NSNS
sector of the perturbative string descriptions and there is no coupling
interpretation for compact dimensions (thus avoiding problems with
strong coupling extrapolations). 

Perhaps the most important aspect of starting in ten-dimensions is
that we have at our disposal the full quantum theory (as opposed to
its low-energy effective field theory limit in the eleven-dimensional
case). Since the \( M^{9}\times S^{1}_{R,B} \) is flat (though globally
nontrivial), string quantization on this background is straightforward.
We can always reduce the theory to nine-dimensions to obtain the corresponding
curved NSNS Melvin backgrounds, but for our purposes the spectrum
calculation in ten-dimensions will suffice.

There are numerous supersymmetric ten-dimensional starting points.
We will first briefly present the ten-dimensional version of the twisted
circle semi-classical instability which parallels section \ref{sec:TwistedCircle}.
This analysis will apply to any perturbative string theory compactified
on a twisted circle\footnote{%
The semiclassical instability arises in the gravity sector which is
common to all five perturbative string theories.
}. We will then move on to a case by case analysis of the small \( R \)
limit. Type IIA/B exhibit similar behavior as do the two heterotic
theories. Type I we discuss on its own.

\subsubsection{Twisted Circle \protect\( M^{9}\times S^{1}_{R,B}\protect \)\label{sec:10twistedcircle}}

Our discussion of the semi-classical instability of twisted circle
compactifications in 11D carries over to this case with little change.
We will quickly highlight the results. The ten-dimensional geometry
is flat with the nontrivial identifications

\begin{eqnarray}
x_{9} & \sim  & x_{9}+2\pi n_{1}R\label{10twistedidentifications} \\
\phi  & \sim  & \phi +2\pi n_{1}BR+2\pi n_{2}.\nonumber 
\end{eqnarray}
The ten-dimensional Euclidean Kerr bounce solution is given by

\begin{eqnarray}
ds_{10}^{2}=(1-\frac{\mu }{r^{5}\Sigma })d\tau ^{2}-\frac{2\mu \alpha sin^{2}\theta }{r^{5}\Sigma }d\tau d\phi +\frac{\Sigma }{r^{2}-\alpha ^{2}-\mu r^{-5}}dr^{2}+\Sigma d\theta ^{2} &  & \label{10Kerrbounce} \\
+\frac{sin^{2}\theta }{\Sigma }[(r^{2}-\alpha ^{2})\Sigma -\frac{\mu }{r^{5}}\alpha ^{2}sin^{2}\theta ]d\phi ^{2}+r^{2}cos^{2}\theta d\Omega _{6}. &  & \nonumber 
\end{eqnarray}
To match this bounce solution to the twisted circle background the
black hole mass and angular momentum parameters \( (\mu ,\alpha ) \)
must satisfy\begin{eqnarray}
R & = & \frac{2\mu }{7r_{H}^{6}-5\alpha ^{2}r_{H}^{4}}\label{10matching} \\
B & = & \frac{\alpha r_{H}^{5}}{\mu }-\frac{\alpha }{|\alpha |R}\nonumber 
\end{eqnarray}
where the horizon radius \( r_{H} \) now satisfies \begin{equation}
\label{10horizon}
r_{H}^{2}=\alpha ^{2}+\frac{\mu }{r_{H}^{5}}.
\end{equation}
The decay rate is given by\begin{equation}
\label{10rate}
\Gamma \sim e^{-\frac{15\pi ^{4}\mu R}{112G_{10}}}
\end{equation}

The post decay evolution of these ten-dimensional theories may be
addressed by repeating the analysis of section \ref{sec:TwistedCircle}
everywhere replacing \( d\Omega _{6}\to d\Omega _{5} \). The picture
is that of an expanding bubble of nothing with surface isometry group
\( SO(2)\times SO(6) \). Two aspects of the semi-classical instability
are important for the discussion in the next section. First of all
the analytically continued bounce solution (bubble) constructed in
section \ref{sec:TwistedCircle} is centered about \( \rho =0 \)
in the plane defining the twist parameter \( B. \) It is difficult
to imagine an off-axis bounce (centered around \( \rho \neq 0 \))
having the correct asymptotic form to be glued into the decaying spacetime,
i.e. \( SO(2) \) isometry defined about \( \rho =0 \). In addition
the pre-decay geometry is translationally invariant along the hyperplane,
so one would expect the semi-classical decay to proceed by nucleating
bubbles with a uniform distribution along the \( \rho =0 \) hyperplane.
These bubbles will expand off the hyperplane eventually affecting
the geometry at all points in the spacetime. For the critically twisted
case \( B=\frac{1}{R} \) the identifications (\ref{10twistedidentifications})
act trivially on the spacetime and the \( U(1)\times SO(6,1)\times SO(2) \)
isometry is restored to a full \( U(1)\times SO(8,1) \). The \( \rho =0 \)
hyperplane is no longer distinguished and so the geometry will decay
by production of spherically symmetric bubbles nucleated throughout
the spacetime.

\subsubsection{Type IIA/B\label{sec:TypeIIA/B} }

The \( R<\sqrt{\alpha '} \) limit of the twisted circle compactification
works out very nicely for the Type IIA/B starting points. The spectra
of these theories has been analyzed in detail in \cite{Russo:2001tf, Russo:1996ik}
and we will recount only a few important aspects of their results.
Our purpose is to compare these results with the semiclassical instability
described above offering support for our conjecture that these instabilites
are related. 

For |\( B|\neq 0,\frac{1}{R} \) the \( 9 \)+\( 1 \)-dimensional
Lorentz invariance of uncompactified theory is broken to \( 6 \)+\( 1 \)-dimensional
Lorentz invariance by the twisted compactification. These theories
contain tachyonic states in the winding sectors for \( R<2\alpha '|B| \).
Combining this with the limited range of the twist parameter \( 0\leq |B|\leq \frac{1}{R} \),
we see that the largest value of \( R \) for which the theory is
tachyonic is \( R=\sqrt{2\alpha '} \) which occurs for the critical
twist \( |B|=\frac{1}{R} \). When \( B=0 \) the theory is supersymmetric
and there are of course no tachyonic modes. For any \( |B|\neq 0 \)
the lowest mass state is a \( (w=1) \) winding mode in the NS+NS+
sector with a negative mass shift due to its angular momentum in the
\( \phi  \)-plane. This arises from a gyromagnetic interaction term
in the string Hamiltonian of the form \begin{equation}
\label{tachyonicmassterm}
-\frac{2BRw}{\alpha '}(\widehat{J}_{R}-\widehat{J}_{L})
\end{equation}
where \( \widehat{J}_{R},\widehat{J}_{L} \) are the angular momentum
operators in the \( \phi  \)-plane. In terms of world-sheet oscillator
excitations it is the same tensor fluctuation that in flat space gives
rise to the graviton. 

For \( |B|\neq 0,\frac{1}{R} \) the winding states (closing only
up to \( n_{1}=w \) in (\ref{10twistedidentifications})) must not
only stretch over the circle, but must also stretch to accomodate
the arc-length subtended by \( 2\pi wBR \). This clearly depends
on the distance \( \rho  \) from the hyperplane about which \( \phi  \)
is defined giving a winding energy contribution to the string mass
of the form\begin{equation}
\label{twistedwindingenergy}
\delta m^{2}=\frac{w^{2}R^{2}}{\alpha '^{2}}(1+\rho ^{2}B^{2}).
\end{equation}
 The tachyonic states in the theory are necessarily winding states
and for \( B\neq 0 \) it is clear that any finite negative mass contribution
will be cancelled for sufficiently large values of \( \rho  \). The
tachyonic states are thus effectively localized about \( \rho =0 \).
This fits in nicely with our conjecture relating the semi-classical
instability for large \( R \) to the tachyonic instability for small
\( R \). In both cases the decay seed is localized to the distinguished
hyperplane. 

One can go even further and analyze the perturbative spectrum for
the critical case \( |B|=\frac{1}{R} \). Naively the argument based
on (\ref{twistedwindingenergy}) would seem to again imply localization
of twisted states. However a careful treatment of string quantization
reveals that for a critical twist the shift in normal ordering constant
restores the zero mode structure in the \( \phi  \)-plane \cite{Russo:1996ik, Russo:2001tf}.
The tachyons are no longer localized to the \( \rho =0 \) hyperplane
in accord with the delocalization of semi-classical bubble production
for the large radius critically twisted circle. This result is not
suprising insofar as we can consider the critically twisted case in
terms of trivial circle compactification with a reversal of spin structure
on the \( S^{1}_{R} \). One can argue that the \( R\to 0 \) limit
of Type IIA/B on a critically twisted (Scherk-Schwarz) can be described
by Type 0A/B on \( M^{9}\times S^{1}_{2R\to 0} \) which are better
described by the T-dual Type0B/A theories on \( M^{10} \).\footnote{%
These arguments again rely on interpolating orbifolds.
} This leads one to pose the following question. Suppose we start with
Type 0A string theory on \( M^{10} \) which has its usual flat-space
tachyon. We wish to connect this tachyon to a semi-classical instability.
There appear to be two candidates. Either M-theory on a critically
twisted circle or Type IIB string theory on a critically twisted circle
of vanishing radius. Though both instabilities lead to the annihilation
of spacetime, the first proceeds via an eleven-dimensional bubble
geometry while the latter proceeds via a ten-dimensional bubble geometry.
This essentially becomes a question of limits. Though \( M^{10} \)
resembles in many ways \( M^{9}\times S^{1}_{R\to \infty } \) there
are global distinctions (for example quantization conditions). If
we are interested in the Type 0A tachyon in strictly \( M^{10} \)
then we should identify it with the M-theory instability since the
limiting theory is fully ten-dimensional (the compactification radius
playing the role of the coupling). The Scherk-Schwarz compactification
of IIB only approaches 0A on \( M^{10} \) as \( M^{9}\times S^{1}_{R\to \infty } \).
However one may still suppose that a similar ambiguity would hold
for Type 0A on \( M^{9}\times S^{1}_{R} \) where \( R \) is finite
and nonzero. In this case either M-theory on \( M^{9}\times S_{R^{11}}^{SS}\times S^{1}_{R} \)
or IIB on \( M^{9}\times S^{SS}_{\frac{\alpha '}{2R}} \) would seem
to work. However the latter does not T-dualize to Type 0A on \( M^{9}\times S^{1}_{R} \),
but rather to Type 0A on \( M^{9}\times \widetilde{S}^{SS}_{\frac{\alpha '}{2R}} \).\footnote{%
The dual Scherk-Schwarz circle may be formally written as \( \frac{S^{1}_{R}}{(-1)^{f_{R}}\times S} \)where
\( f_{R} \) is the right moving world-sheet fermion number \cite{Bergman:1999km}.
} Again the appropriate instability is the M-theory one.

\subsubsection{Heterotic \protect\( SO(32)/E8\times E8\protect \)\label{sec:Heterotics}}

The twisted circle semi-classical instability is generic. Any theory
with a gravity sector formulated on this background will decay by
nucleating (possibly deformed) bubbles of nothing. If the twisted
circle radius is small enough the semi-classical calculation is no
longer valid and we believe the same underlying instability should
emerge in whatever description becomes appropriate. We can push this
further and say that since the semi-classical instability is generic
then, at least for twisted circle compactifications, the corresponding
tachyonic instabilities should be generic. For theories with gauge
degrees of freedom this can be a nontrivial issue. 

Consider taking either of the supersymmetric heterotic theories and
compactifying on a twisted circle. The nontrivial boundary conditions
for arbitrary \( B \) will affect both bosons and fermions winding
around the compact circle, but should leave massless the nonwinding
gauge bosons of the heterotic theory. The gauge symmetry should thus
remain unbroken. It would be natural then to expect that the tachyonic
instability will be neutral under the heterotic gauge symmetry. We
expect the spectrum of heterotic strings on a twisted circle to be
very similar to the spectrum of Type II strings on a twisted circle
barring the usual differences between the spectra of the free theories.
A cursory investigation of the spectrum of these theories in \cite{Russo:1996ik}
supports this picture. In particular the negative mass contributions
attributable to the gyromagnetic interaction term renders the lowest
NS+NS+ states tachyonic. This gauge neutral tachyonic state is exactly
what we expect from our conjecture relating it to the semi-classical
instability. The story changes considerably if we include a Wilson
line along the twisted circle which breaks the heterotic gauge symmetry
to some subgroup. In this case the semi-classical instability is associated
with a gauge symmetry breaking compactification and we expect the
corresponding tachyons to transform nontrivially under the unbroken
gauge group. The details of the spectrum are currently under investigation.

The critical case without a Wilson line poses an interesting problem
for the heterotic strings. Consider setting \( |B|=\frac{1}{R} \)
and sending the compactification radius to zero. By the interpolating
orbifold argument the resulting description can be described in terms
of a nonsupersymmetric string theory on \( M^{10} \) that is the
T-dual of the result of twisting the orginal theory by \( (-1)^{F_{s}} \).
Surveying the table of consistent nonsupersymmetric heterotic strings
in section \ref{subsec:ShrinkInterval} we find that there is no flat
space theory with gauge group \( E8\times E8 \). In addition the
theory with gauge group \( SO(32) \) exhibits a tachyon which transforms
nontrivially under the gauge group in contradiction to the expectations
outlined above. The resolution of this puzzle is what makes the heterotic
case so interesting. The two supersymmetric heterotic string theories
are actually invariant under a twist by \( (-1)^{F_{s}} \) as pointed
out in \cite{Dixon:1986iz}. Combined with the self-duality of these
theories under T-duality, the result is that the \( R\to 0 \) limit
of a critically twisted compactification of these theories returns
the original theory on \( M^{9}\times S^{1}_{\alpha '/R\to \infty } \).
In this case the association is not between a semi-classical instability
on one hand and a tachyonic instability on the other, but rather between
semi-classical instabilities both large and small compactification
radii. This case represents the single exception we have found to
the strong form of our conjecture. If we include a Wilson line along
the critically twisted circle, then the \( R\to 0 \) limit can be
described by the T-duals of the nonsupersymmetric heterotic theories
and the delocalized semi-classical instabilities descend to the bulk
tachyons. 

Our conjecture then implies that the fate of these theories after
condensation of the closed string tachyon involves the annihilation
of spacetime. 

This poses a resolution to an issue raised by Suyama \cite{Suyama:2001gd, Suyama:2001ik}.
Working under the assumption that condensation of the tachyon in the
nonsupersymmetric heterotic theories takes the theories to a stable
supersymmetric background the only choices for the endpoint involve
the gauge symmetries \( SO(32),E8\times E8 \). It was pointed out
that condensation of a tachyon transforming nontrivially under a gauge
symmetry should reduce the rank of the gauge group. However, the ranks
of the gauge symmetries for the nonsupersymmetric heterotic strings
are already as small or smaller than in the two supersymmetric cases.
Our conclusion, i.e. that condensation of the tachyon leads to an
annihilation of the spacetime, avoids this puzzle entirely. 

A notable exception is the \( H_{SO(16)\times SO(16)} \) theory which
exhibits no tachyonic instability. This theory may be obtained as
the T-dual of a Scherk-Schwarz orbifold of either \( H^{susy}_{SO(32)} \)
or \( H^{susy}_{E8\times E8} \) with appropriate Wilson lines. One
expects a semiclassical instability towards spacetime annihilation,
however in this case it is unclear to what the semi-classical instability
descends. However this case is unique in that this particular Scherk-Schwarz
compactification generates a positive cosmological constant. For all
other Scherk-Schwarz compactifications the cosmological constant is
negative driving the theory towards compactification and thus towards
the tachyonic regime. In this case the positive cosmological constant
generates a potential which pushes the theory towards decompactification
thereby restoring the supersymmetric background with which we began.

\section{Conclusions}

Our aim in this paper has largely been twofold. On the one hand we
have taken seriously the idea that semi-classical gravitational instabilities
in supersymmetry breaking compactifications may in certain limits
reduce to perturbative instabilities signalled by the appearance of
a closed string tachyon. In making this identification it is then
natural to identify the endpoint of condensation of the tachyon with
the endpoint of the semi-classical instability. In every case that
we have considered the endpoint involves an annihilation of the metric
degrees of freedom. We have further made a case for the naturalness
of such a catastrophic fate by comparing these theories to those theories
exhibiting open string tachyons for which extensive evidence has been
presented. In both cases the corresponding degrees of freedom are
annihilated.

On the other hand we have used this connection between semi-classical
instabilities and tachyons to explore a possible web of dualities
involving nonsupersymetric string theories. In particular the eleven-dimensional
origins of many nonsupersymmetric ten-dimensional string backgrounds
has been conjectured and the overall picture appears to hang together
quite nicely. Our discussion of the limits of the Fabinger-Horava
theory constitute to our knowledge the first attempt to extend the
0A/M-theory relation of Bergman and Gaberdiel \cite{Bergman:1999km}
to the heterotic theories.\footnote{%
In an earlier version of this paper we discussed a simple Scherk-Schwarz
compactification of the Horava-Witten background. After several discussions
with Oren Bergman we feel that this background deserves more investigation
and so we leave this analysis to future work \cite{Bergman:Inprogress}.
}

A by product of our arguments is that Scherk-Schwarz compactification
is not a very useful tool for constructing phenomenological SUSY breaking
theories. In this the usual problem has been that the radius \( R \)
of the compactification circle would tend to zero because of the potential
that developes at one loop (and higher) \cite{Rohm:1984aq}.\footnote{%
We wish to thank Oren Bergman for pointing out an error in the first
of this paper where we stated that the circle tends to decompactify.
Though the restoration of supersymmetry seems natural in this context,
the sign of the potential found by Rohm does in fact indicate that
the circle is driven towards smaller radius.
} Thus the system approaches the tachyonic regime. However one might
imagine that this modulus is stabilized either by classical flux terms
or by some non-perturbative quantum effect. One would of course want
this stabilization to occur at some \( R>\sqrt{\alpha '} \) , in
order to avoid having a tachyon (and also usually to get smaller than
string scale SUSY breaking). However at such radii the semi-classical
instability of Witten that we discussed extensively in this paper
takes over.\footnote{%
An issue associated with this is the effect of a conserved flux on
the spacetime annihilation picture. At present this is unclear. We
thank Yonatan Zunger for pointing this out. 
} It may be possible of course that the fluxes (or quantum effects)
are such as to stabilize the radius at a large enough value such that
the semi-classical decay lifetime is larger than the age of the universe,
but this strikes us as being somewhat unnatural.

There are a number of outstanding issues associated with our conclusions.
First of all a quantitative description of the condensation of closed
string tachyons in a vein similar to the open string case would put
all of these speculations on a much firmer footing. The bounce solutions
for Scherk-Schwarz orbifolds with Wilson lines are under consideration.
These might shed light on a nonperturbative framework for the {}``other{}''
nonsupersymmetric heterotic theories.\footnote{%
The equivalence of supersymmetric heterotic strings on SUSY-breaking
Melvin backgrounds to nonsupersymmetric heterotic strings has been
thoroughly investigated in \cite{Suyama:2001gd}\cite{Suyama:2001ik}\cite{Suyama:2001ne}\cite{Suyama:2001bn}.
} Subtleties associated with the zero-radius limit of Scherk-Schwarz
compactifications are unresolved but under current investigation \cite{Bergman:Inprogress}.

\section{Acknowledgements}

One of us wishes to thank the hospitality of the ITP Santa Barbara
and the organizers of the M-theory work shop, and Gary Horowitz, Joe
Polchinski and Eva Siverstein for discussions during the initial stages
of this work. We also wish to thank Steve Giddings, Michael Gutperle,
Yonatan Zunger, and Brookie Williams for discussions. Oren Bergman
provided a number of corrections, references, and useful discussion
during the revision of this work. This research was supported in part
by the United States Department of Energy under grant DE-FG02-91-ER-40672.

\section*{Appendix}

\newpage
\bibliographystyle{apsrevgoodtitle}
\bibliography{CSTASI}

\end{document}